\documentclass[superscriptaddress,aps,letterpaper,10pt,twocolumn,floats,showpacs,amsmath,amsfonts,amssymb,prb]{revtex4-2}

\usepackage{graphicx}
\usepackage[colorlinks=true,citecolor=blue]{hyperref}
\usepackage[caption=false]{subfig}
\usepackage{pstricks}
\usepackage{pst-node}
\usepackage{amsmath}
\usepackage{amsbsy}
\usepackage{verbatim}
\usepackage{dsfont}
\begin{document}


\title{Finite-frequency noise in non-Markovian systems: A Markovian embedding approach}

\author{Krzysztof Ptaszy\'{n}ski}
\affiliation{Institute of Molecular Physics, Polish Academy of Sciences, Mariana Smoluchowskiego 17, 60-179 Pozna\'{n}, Poland}
\email{krzysztof.ptaszynski@ifmpan.poznan.pl}

\date{\today}

\begin{abstract}
I present an approach to calculate the finite-frequency quantum noise in systems strongly coupled to structured reservoirs based on 
the Markovian embedding. This technique consists of mapping of the non-Markovian dynamics of the original system onto the Markovian dynamics of the extended supersystem. The applicability of this method is demonstrated on a single electronic level coupled to a reservoir with an energy-dependent density of states. This model can be mapped onto an equivalent double-level system whose dynamics is governed by an exact Markovian master equation. It is demonstrated that the presented approach provides a way to describe genuinely quantum effects, such as the asymmetry of the absorption and the emission noise, as well as to deal with non-Markovian effects and intra-system interactions on equal footing.
\end{abstract}

\maketitle

\section{Introduction}
Current fluctuations have been well established as an important tool for the characterization of transport mechanism in nano-scale conductors~\cite{blanter2000}. In particular, a rich information about the microscopic dynamics of nanoelectronic systems is provided by the finite-frequency noise which characterizes the temporal correlations of the current. For a multiterminal device attached to several leads the nonsymmetrized current noise is defined as~\cite{engel2003, zamoum2016}
\begin{align} \label{noise}
S_{\alpha \beta}(\omega)=\int_{-\infty}^{\infty} dt e^{-i \omega t}\langle \Delta \hat{I}_\alpha(t) \Delta \hat{I}_\beta(0) \rangle,
\end{align}
where $\hat{I}_\alpha$ is the operator (in the Heisenberg picture) describing the current to the lead $\alpha$, $\langle ... \rangle$ denotes the steady-state average and $\Delta \hat{I}_\alpha = \hat{I}_\alpha - \langle \hat{I}_\alpha \rangle$ is the deviation from the average current. For $\alpha = \beta$ ($\alpha \neq \beta$) this quantity describes the correlation of the current flowing through a single junction (different junctions), and is thus referred as the autocorrelation noise (cross-correlation noise). As the name suggests, the nonsymmetrized noise may be asymmetric with respect to the frequency [$S_{\alpha \beta}(\omega) \neq S_{\alpha \beta}(-\omega)$], which is due to the non-commutation of the current operators for different times~\cite{engel2003, zamoum2016}.

The finite-frequency noise can, on the one hand, reveal the characteristic time scales of the system associated, e.g., with the relaxation processes~\cite{bulka1999, michalek2020} or the internal coherent oscillations~\cite{braun2006, luo2007, lambert2008, dong2008, jin2013, droste2015, xue2018, kirton2012, flindt2008, marcos2011}. On the other hand, there is a connection between the noise spectrum and the energy scales of the system since the current autocorrelation noise $S_{\alpha \alpha}(\omega)$ can be related to the emission ($\omega>0$) or the absorption ($\omega<0$) of photons with an energy $\hbar \omega$ which can be detected by an appropriate quantum detector~\cite{lesovik1997, aguado2000, gavish2000}. Indeed, detection of the nonsymmetrized quantum noise~\cite{deblock2003, billangeon2008, basset2010} or its emission part~\cite{basset2012, delagrange2018, fevrier2018} is now well established experimentally. From the theoretical side, it was demonstrated that different contributions to the noise spectrum can be related to different microscopic inelestic scattering (energy transfer) processes~\cite{zamoum2016, crepieux2018}. It should be emphasized that the information about the asymmetry of the absorption and the emission spectra is missed by a symmetrized noise $S^\text{sym}_{\alpha \alpha}(\omega)={[S_{\alpha \alpha}(\omega)+S_{\alpha \alpha}(-\omega)]/2}$ which is often considered in the literature~\cite{engel2003, zamoum2016}.

There are several theoretical approaches to calculate the finite-frequency noise. The methods based on non-equilibrium Green's functions provide an exact characterization of the noise in noninteracting systems~\cite{zamoum2016}. However, for interacting systems the approximate perturbative approaches are required~\cite{moca2011, orth2012, crepieux2018, stadler2018}. The methods based on master equations enable one, one the other hand, to deal with intra-system interactions in an exact way, whereas the coupling to the environment is treated perturbatively. In particular, most studies focus on the Markovian regime in which the memory effects associated with the system-environment correlations are neglected~\cite{braun2006, luo2007, lambert2008, dong2008, jin2013, droste2015, xue2018, flindt2004, kirton2012, bulka1999, michalek2020}. The Markovian master equations can be derived by assuming a weak coupling to the environment~\cite{breuer2002, schaller2014}. Alternatively, as shown by Gurvitz and Prager~\cite{gurvitz1996, gurvitz1998}, for fermionic systems an exact Markovian master equation, applicable also for the strong coupling, can be derived in the limit of infinite voltage bias. However, this method is only valid for an energy-independent coupling to the environment, whereas the energy dependence of the coupling may lead to strongly non-Markovian effects~\cite{zedler2009}. Furthermore, the Markov approximation misses the quantum nature of the noise by providing only a symmetrized spectrum~\cite{engel2003}. Whereas one can generalize the master equation approach beyond the Markovian regime~\cite{flindt2008, marcos2011}, this may still fail to correctly describe the noise when the dynamics is strongly non-Markovian~\cite{zedler2009}.

Here I show that non-Markovian effects and intra-system interactions can be dealt with on equal footing using the Markovian embedding. Within this approach the original system $S$ is embedded in the auxiliary system $A$, such that the joint supersystem $SA$ undergoes a Markovian dynamics~\cite{arrigoni2013, dorda2014, woods2014, tamascelli2018, chen2019, chen2019b}. In particular, the paper analyzes the finite-frequency noise in a single electronic level (e.g, in a quantum dot) weakly coupled to the source lead while strongly coupled to the drain lead with an energy-dependent spectral density. This model can be exactly mapped onto an equivalent double-level model whose dynamics is Markovian. It is shown that the non-Markovian character of the dynamics can be revealed by pronounced peaks or dips in the noise spectrum associated with the coherent oscillations between the level and the drain lead, as well as by the asymmetry of the noise spectrum. It is also demonstrated that the on-site Coulomb interaction can, on the one hand, strongly enhance the emission peak, which is due to the inelastic nature of the tunneling from a doubly-occupied level to the drain lead; on the other hand, it can induce a switching between the transport channels with a high and a low conductance, leading to the enhancement of the zero-frequency noise to super-Poissonian values. 

The paper is organized as follows. Sec.~\ref{sec:embgen} discusses a general idea of the Markovian embedding while Sec.~\ref{sec:rcm} focuses on a particular form of this approach based on the reaction coordinate mapping. Sec.~\ref{sec:noise} presents the methods used to calculate the finite-frequency noise in systems described by Markovian master equations. Secs.~\ref{sec:nrl} and~\ref{sec:and} demonstrate the results obtained for the noninteracting and the interacting single-level model, respectively. Finally, Sec.~\ref{sec:concl} draws the conclusions. The appendices~\ref{sec:repr} and~\ref{sec:negf} contain some technical details concerning the calculation of the noise.

\section{Markovian embedding} \label{sec:emb}
\subsection{General idea} \label{sec:embgen}
To introduce the Markovian embedding approach, let me first remind some basic ideas concerning the open system dynamics. A generic open quantum system is described by the Hamiltonian~\cite{breuer2002, schaller2014}
\begin{align} \label{hamop}
\hat{H}_{SE}=\hat{H}_S+\hat{H}_E+\hat{V}_{SE},
\end{align}
where $\hat{H}_S$, $\hat{H}_E$, $\hat{V}_{SE}$ are the Hamiltonians of the system, environment and the interaction between them, respectively. The evolution of the joint state of the system and the environment $\rho_{SE}$ is described by the von Neumann equation
\begin{align}
\frac{d}{dt} \rho_{SE}(t) =-\frac{i}{\hbar} \left[ \hat{H}_{SE}, \rho_{SE}(t) \right].
\end{align}
Due to the large size of the bath, the exact dynamics of $\rho_{SE}(t)$ is usually intractable. However, one is mostly interested in the dynamics of the reduced state of the system
\begin{align}
\rho_S(t)=\text{Tr}_E \left[ \rho_{SE}(t)\right],
\end{align}
where $\text{Tr}_E$ is the partial trace over the state of the environment. Different theoretical techniques (usually approximate) to describe the reduced dynamics have been developed. In certain regimes (e.g., for a weak coupling to the environment) one can apply the Born-Markov approximation which neglects the memory effects related to the correlation between the system and the environment~\cite{breuer2002, schaller2014}. In such a case the reduced dynamics can be described using a time local master equation
\begin{align}
\frac{d}{dt} \rho_S(t) = \mathcal{L}_S \rho_S(t),
\end{align}
where superoperator $\mathcal{L}_S$ is referred to as the Liouvillian. 

The Born-Markov approximation may fail, for example, when the coupling to the bath is strong or strongly energy-dependent. To deal with this problem the Markovian embedding technique has been developed~\cite{arrigoni2013, dorda2014, woods2014, tamascelli2018, chen2019, chen2019b}. Within this approach the system $S$ is placed into contact with the auxiliary system $A$, such that dynamics of the joint supersystem $SA$ can be described using the Markovian master equation
\begin{align} \label{mastereqemb}
\frac{d}{dt} \rho_{SA}(t) = \mathcal{L}_{SA} \rho_{SA}(t).
\end{align}
The state of the system can be then determined as
\begin{align}
\rho_S(t)=\text{Tr}_A \left[ \rho_{SA}(t)\right],
\end{align}
where $\text{Tr}_A$ is the partial trace over the state of the auxiliary system. It was rigorously proven that for systems with bosonic~\cite{tamascelli2018} or fermionic~\cite{chen2019} Gaussian environments an exact Markovian embedding is in principle always applicable, and different theoretical techniques methods to realize this idea in practice have been developed~\cite{arrigoni2013, dorda2014, woods2014, chen2019b}. 

%
\begin{figure} 
	\centering
	\includegraphics[width=0.9\linewidth]{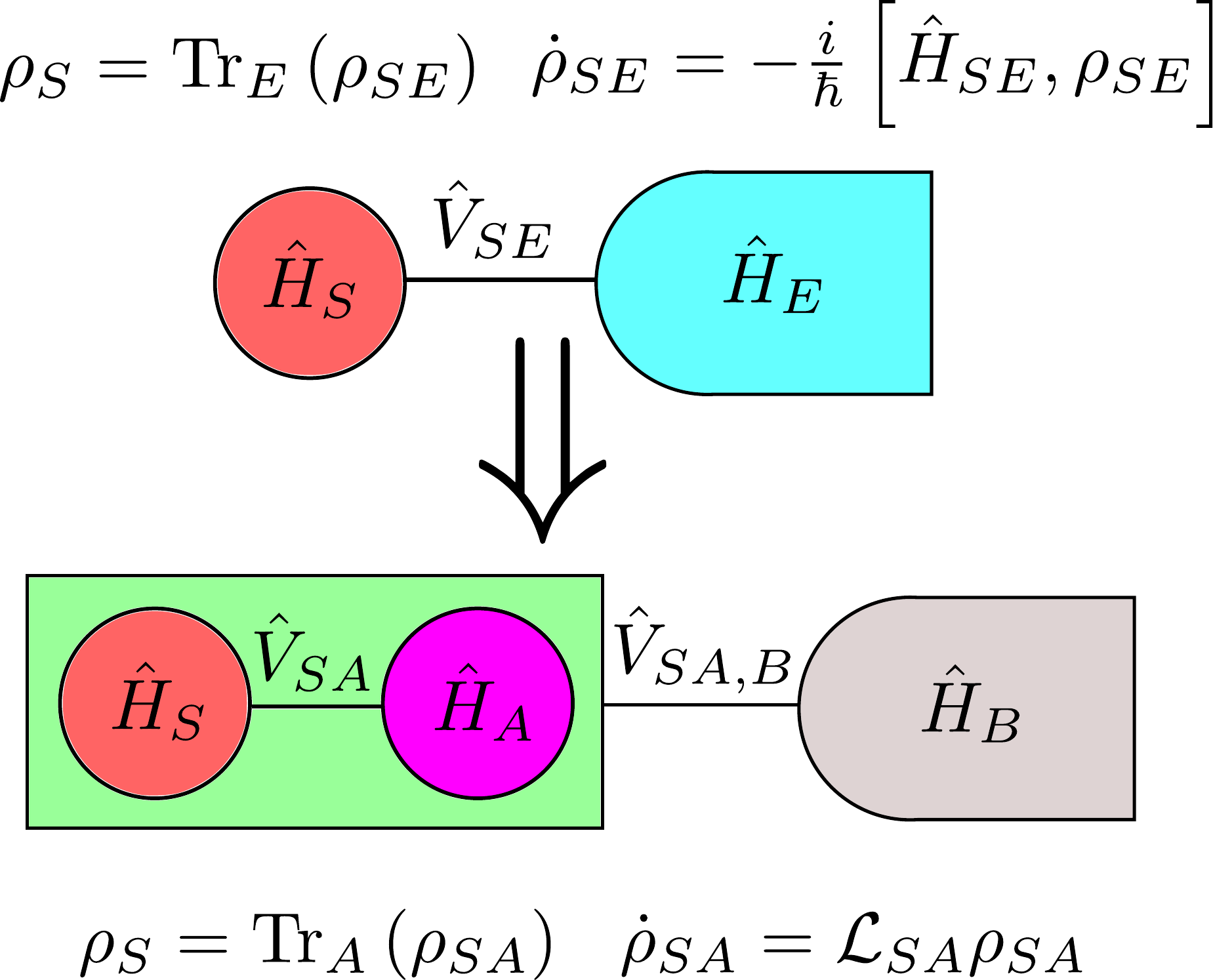} 
	\caption{A schematic representation of the Markovian embedding. The original open system $S$ attached to the environment $E$ is placed into contact with the auxiliary system $A$ such that the joint supersystem $SA$, attached to the residual bath $B$, undergoes a Markovian dynamics.}
	\label{fig:markemb}
\end{figure}
%
From a physical point of view, in many situation the Markovian embedding consists of the transformation of the original open system Hamiltonian~\eqref{hamop} into the auxiliary form~\cite{woods2014, tamascelli2018, chen2019}
\begin{align} \label{hamaux}
\hat{H}_{SE}^\text{aux}=\hat{H}_S+\hat{H}_A+\hat{V}_{SA}+\hat{H}_B+\hat{V}_{SA,B},
\end{align}
where $\hat{H}_A$ is the Hamiltonian of the auxiliary system, $\hat{V}_{SA}$ describes the coupling between the original and the auxiliary system, $\hat{H}_B$ is the Hamiltonian of the residual bath and $\hat{V}_{SA,B}$ corresponds to the coupling between the supersystem $SA$ and the residual bath (see a schematic representation in Fig.~\ref{fig:markemb}). One may then apply the standard techniques to derive the Markovian master equation describing the dynamics of the supersystem $SA$. In the next section I will focus on a particular form of such a transformation, referred to as the reaction coordinate mapping.

\subsection{Reaction coordinate mapping} \label{sec:rcm}
The reaction coordinate mapping~\cite{smith2014, strasberg2016, strasberg2018, nazir2018} is a method used to obtain the auxiliary Hamiltonian~\eqref{hamaux} which is applicable to systems coupled linearly to continuous quadratic bosonic or fermionic baths. Here I focus on the fermionic case; the formalism applicable to bosonic baths is presented in Refs.~\cite{smith2014, strasberg2016, nazir2018}. In particular, a single bath tunnel-coupled to a single fermionic site in the system will now be considered; a generalization to multiple baths and multiple sites is straightforward. The Hamiltonians $\hat{H}_E$ and $\hat{V}_{SE}$ read then as~\cite{strasberg2018, nazir2018}
\begin{align}
\hat{H}_{E}&=\sum_k \epsilon_k c_k^\dagger c_k, \\
\hat{V}_{SE}&=\sum_{k} \left( t_{k} C^\dagger c_k+\text{h.c.} \right),
\end{align}
where $c_k^\dagger$ ($c_k$) is the (creation) annihilation operator of the fermion on the site $k$ in the environment, whereas $C^\dagger$ and $C$ are the corresponding system operators. The coupling to the bath is characterized by a spectral density
\begin{align}
\Gamma (\epsilon)=2 \pi \sum_{k} |t_{k}|^2 \delta(\epsilon-\epsilon_k).
\end{align}
%
\begin{figure} 
	\centering
	\includegraphics[width=0.9\linewidth]{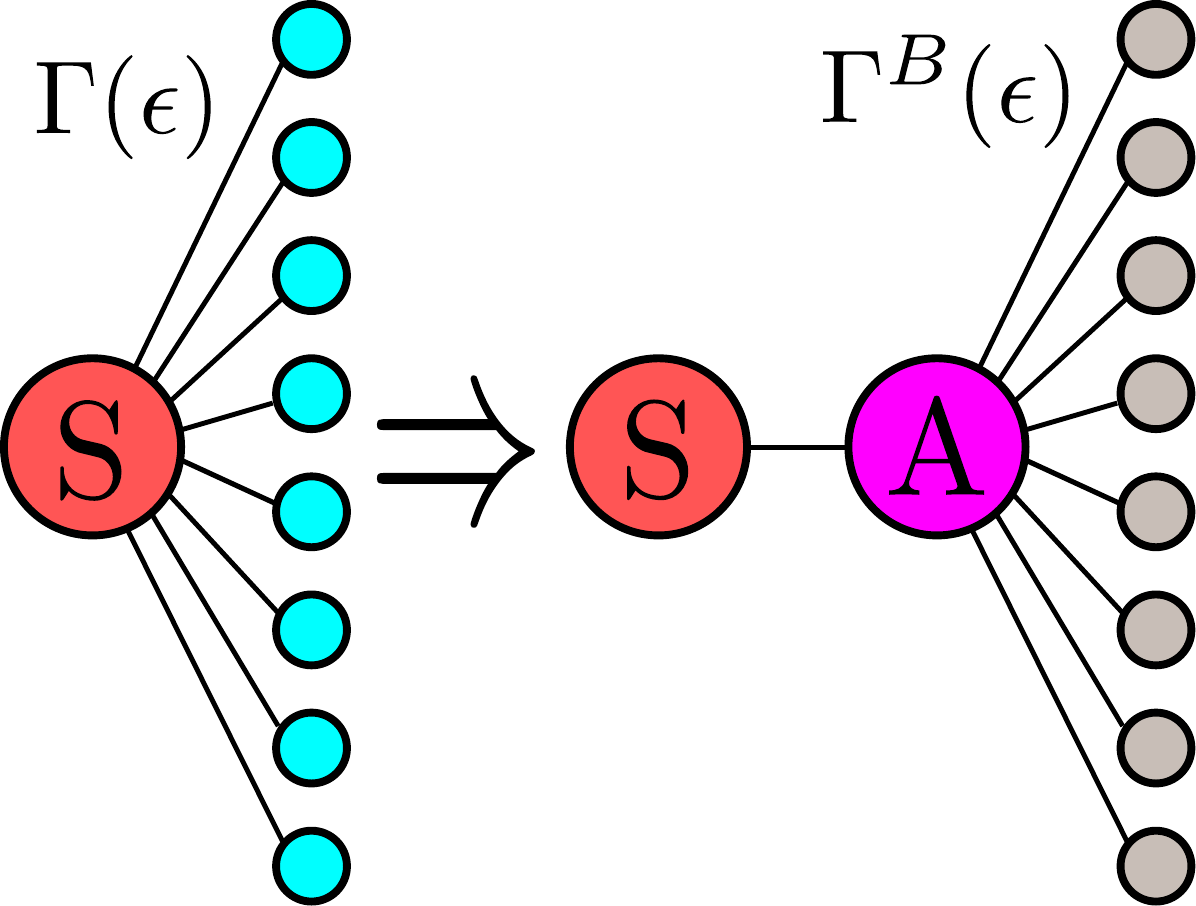} 
	\caption{A schematic representation of the reaction coordinate mapping. The original system connected to a fermionic bath with a spectral density $\Gamma(\epsilon)$ is transformed into the system attached through the auxiliary site $A$ to a residual fermionic bath with a spectral density $\Gamma^B(\epsilon)$.}
	\label{fig:rcm}
\end{figure}
%
The auxiliary system consists then of a single fermionic site tunnel coupled to the original system, i.e,
\begin{align}
\hat{H}_{A}&=\varepsilon D^\dagger D, \\
\hat{V}_{SA}&=\lambda C^\dagger D+\text{h.c.},
\end{align}
where $D^\dagger$ ($D$) is the creation (annihilation) operator of the fermion on the auxiliary site (see a schematic representation in Fig.~\ref{fig:rcm}). The energy of the auxiliary site $\varepsilon$ and the tunnel coupling $\lambda$ are given by the equations
\begin{align}
\lambda^2&=\frac{1}{2\pi} \int_{-\infty}^{\infty} \Gamma(\epsilon) d\epsilon , \\
\varepsilon& = \frac{1}{2\pi \lambda^2} \int_{-\infty}^{\infty} \epsilon \Gamma(\epsilon) d\epsilon.
\end{align}
The auxiliary site is then coupled to the residual fermionic bath $B$, which is characterized by the Hamiltonians
\begin{align}
\hat{H}_B&=\sum_{k} \varepsilon_k d_k^\dagger d_k, \\
\hat{V}_{SA,B}&= \sum_{ik} \left( T_{k} D^\dagger d_k+\text{h.c.} \right),
\end{align}
where $d^\dagger_k$ ($d_k$) is the creation (annihilation) operator of the fermion on the site $k$ in the residual bath. The coupling of the auxiliary site to the residual bath is characterized by a modified spectral density
\begin{align}
\Gamma^B (\epsilon)=2 \pi \sum_{k} |T_{k}|^2 \delta(\epsilon-\varepsilon_k)
\end{align}
which is given by the equation
\begin{align}
\Gamma^B (\epsilon)=\frac{4 \lambda^2 \Gamma(\epsilon)}{\left[\frac{1}{\pi} \mathcal{P} \int_{-\infty}^{\infty} \frac{\Gamma(\epsilon') d\epsilon'}{\epsilon'-\epsilon} \right]^2+\left[\Gamma(\epsilon) \right]^2},
\end{align}
where $\mathcal{P}$ denotes the principal part of the integral.

\section{Finite frequency noise} \label{sec:noise}
I will now present methods to calculate the finite-frequency noise within the Markovian embedding approach. To this goal, let us first note that the master equation~\eqref{mastereqemb} can be decomposed into two contributions~\cite{breuer2002, schaller2014}:
\begin{align} \label{mastereqd}
\frac{d}{dt} \rho_{SA} = \mathcal{L}_{SA} \rho_{SA} = -\frac{i}{\hbar} \left[ \hat{H}_{SA}^\text{eff}, \rho_{SA} \right]+ \mathcal{D}_{SA} \rho_{SA}.
\end{align}
The first term,
\begin{align}
-\frac{i}{\hbar} \left[ \hat{H}_{SA}^\text{eff}, \rho_{SA} \right],
\end{align}
describes the internal, unitary dynamics of the supersystem $SA$. Here $\hat{H}_{SA}^\text{eff}$ is the effective Hamiltonian; it may differ from the original Hamiltonian $\hat{H}_{SA}$ due to renormalization effects caused by the interaction with the residual bath $B$~\cite{breuer2002, schaller2014}. The second contribution,
\begin{align}
\mathcal{D}_{SA} \rho_{SA},
\end{align}
corresponds to a dissipative dynamics related to the coupling to the residual bath. Here the superoperator $\mathcal{D}_{SA}$ is referred to as the dissipator; it describes the classical jump processes between the states of the supersystem $SA$. Accordingly, the current operators in the Hilbert space of $SA$ may now correspond to:
\begin{itemize}
	\item the currents within the supersystem $SA$, (e.g., between the system $S$ and the auxiliary system $A$), later referred to as the coherent currents,
	\item the currents between the supersystem $SA$ and the residual bath $B$, later referred to as the dissipative currents.
\end{itemize}
This distinction will be further shown to be important. Whereas most studies applying the master equation approach have focused on the noise of dissipative currents~\cite{braun2006, luo2007, lambert2008, dong2008, jin2013, droste2015, xue2018, bulka1999, michalek2020}, fluctuations of the coherent currents have been analyzed in Refs.~\cite{flindt2004, kirton2012}. In this paper both types of currents will be considered within a unified approach.

To shorten the notation from hereon I will denote $I_\alpha(0)=I_\alpha$ and omit the subscript $SA$ in $\mathcal{L}_{SA}$. Furthermore, the paper focuses on the case when the Liouvillian $\mathcal{L}$ is time-independent. Due to the equivalence of the Heisenberg and the Schr\"{o}dinger picture the time-evolved current operator $I_\alpha(t)$ is defined by the relation 
\begin{align}
\text{Tr} \left[I_\alpha(t) \rho_{SA}(0) \right]=  \text{Tr} \left[I_\alpha e^{\mathcal{L} t} \rho_{SA}(0) \right].
\end{align}
This suggests that analogously
\begin{align} \label{curcorprov}
\langle \hat{I}_\alpha (t) \hat{I}_\beta \rangle =\text{Tr} \left[ \hat{I}_\alpha(t) \hat{I}_\beta \rho_{SA}^\text{st} \right]=\text{Tr} \left[ \hat{I}_\alpha e^{\mathcal{L} t} \hat{I}_\beta \rho_{SA}^\text{st} \right],
\end{align}
where $\rho_{SA}^\text{st}$ is the density matrix of the stationary state, i.e., the solution of the equation $\mathcal{L} \rho_{SA}^\text{st}=0$ (for a more rigorous derivation, see Ref.~\cite{flindt2004}). There are, however, two caveats related to a peculiar nature of the dissipative (non-unitary) evolution. First, in contrast to the unitary dynamics, the evolution given by Eq.~\eqref{mastereqd} is irreversible. Therefore, Eq.~\eqref{curcorprov} is only valid for $t>0$. This problem can be dealt with by using the relation $\langle \hat{I}_\alpha (-t) \hat{I}_\beta \rangle=\langle \hat{I}_\beta (t) \hat{I}_\alpha \rangle^*$~\cite{engel2003} which can be proven by going back to the Hilbert space of the system and the environment:
\begin{align} \nonumber
&\langle \hat{I}_\alpha (-t) \hat{I}_\beta \rangle = \text{Tr} \left[ \hat{I}_\alpha (-t) \hat{I}_\beta \rho_{SE} \right]=\text{Tr} \left[ \rho_{SE}  \hat{I}_\beta \hat{I}_\alpha (-t) \right]^* \\
&=\text{Tr} \left[ \hat{I}_\beta (t) \hat{I}_\alpha \rho_{SE} \right]^*=\langle \hat{I}_\beta (t) \hat{I}_\alpha \rangle^*.
\end{align}
Here in the second step the relation $\text{Tr}(\hat{O} \hat{P} \hat{R})=\text{Tr}(\hat{R}^\dagger \hat{P}^\dagger \hat{O}^\dagger)^*$ is applied, whereas in the third step I use the cyclic property of the trace and take the dynamics to be stationary.

Secondly, the dissipator $\mathcal{D}_{SA}$ describes the instantaneous jump processes, whereas a genuine quantum evolution is always time-continuous. As a consequence, a singular correction to the auto-correlations of dissipative currents for $t=0$ is required~\cite{korotkov1994}. It is given by the second term in the equation
\begin{align}
\langle \hat{I}_\alpha (t) \hat{I}_\beta \rangle=\text{Tr} \left[ \hat{I}_\alpha e^{\mathcal{L} t} \hat{I}_\beta \rho_{SA}^\text{st} \right]+\delta^\mathcal{D}_{\alpha \beta} \delta(t) \langle I_\alpha \rangle,
\end{align}
which is valid for $t \geq 0$. Here 
\begin{align}
\langle I_\alpha \rangle=\left[ \hat{I}_\alpha \rho_{SA}^\text{st} \right]
\end{align}
is the mean current to the lead $\alpha$ and $\delta^\mathcal{D}_{\alpha \beta}$ is equal to 1 for the auto-correlation ($\alpha=\beta$) of a dissipative current, while equals 0 otherwise. The singular correction, derived by Korotkov~\cite{korotkov1994}, is referred to as the Schottky term or the self-correlation term~\cite{flindt2004}.

Using the relations above one gets a formula for the finite-frequency nonsymmetrized noise,
\begin{align} \label{noiseform}
&S_{\alpha \beta}(\omega) = \int_{-\infty}^{\infty} dt e^{-i \omega t} \langle \Delta \hat{I}_\alpha (t) \Delta \hat{I}_\beta \rangle \\ \nonumber
&= \int_{0^+}^{\infty} dt \left[ e^{-i \omega t} \langle \Delta \hat{I}_\alpha (t) \Delta \hat{I}_\beta \rangle+e^{i \omega t} \langle \Delta \hat{I}_\beta (t) \Delta \hat{I}_\alpha \rangle^* \right]\\ \nonumber
&+\int_{-\infty}^{\infty} dt e^{-i \omega t} \delta^\mathcal{D}_{\alpha \beta} \delta(t) \langle \hat{I}_\alpha \rangle \\ \nonumber
&=\text{Tr} \left(\hat{I}_\alpha \frac{1}{i \omega \mathds{1}+\mathcal{L}} \hat{I}_\beta \rho_{SA}^{st} \right)+\text{Tr} \left(\hat{I}_\beta \frac{1}{i \omega \mathds{1}+\mathcal{L}} \hat{I}_\alpha \rho^{st}_{SA} \right)^* \\ \nonumber &+\delta^\mathcal{D}_{\alpha \beta} \langle \hat{I}_\alpha \rangle,
\end{align}
where $\mathds{1}$ is the identity operator (cf. a similar formula for the symmetrized noise in Ref.~\cite{flindt2004}); here in the last step the formal Fourier transform of $\exp(\mathcal{L} t)$ is applied~\cite{engel2003, flindt2004}. The practical calculation of the noise is most convenient in the Liouville space representation where the density matrix $\rho^{st}_{SA}$ is represented by a column vector containing both the diagonal and the off-diagonal elements of the density matrix while other operators and superoperators are represented by square matrices~\cite{flindt2004, brandes2008}. Details of this approach are presented in the Appendix~\ref{sec:repr}.

Finally, let us discuss some features of the finite-frequency noise. As follows from Eq.~\eqref{noiseform}, $S_{\alpha \beta}(\omega)=S_{\beta \alpha}(\omega)^*$~\cite{engel2003}. Therefore, as already mentioned, the auto-correlation noise $S_{\alpha \alpha}(\omega)$ is real, and thus corresponds to a physical observable. On the other hand, the nonsymmetrized cross-correlation noise may be not real, and therefore is not an observable quantity~\cite{zamoum2016}. However, as theoretically shown by Creux et al.~\cite{creux2006}, its real part is measurable.

\section{Noninteracting resonant level} \label{sec:nrl}
\subsection{Model}
\begin{figure} 
	\centering
	\includegraphics[width=0.9\linewidth]{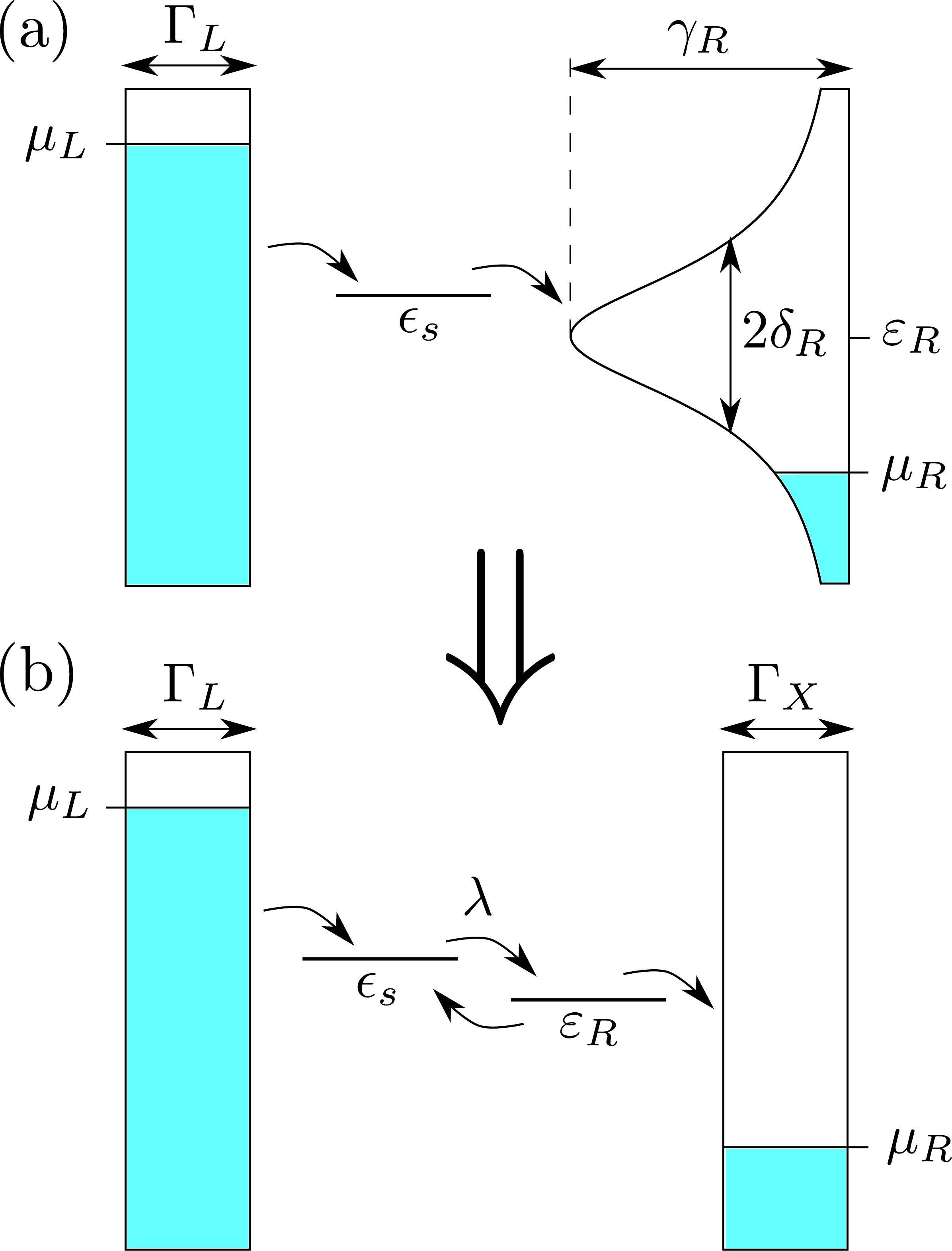} 
	\caption{(a) The original model of a single fermionic level attached to the wide-band left lead and the right lead with a Lorentzian spectral density. (b) The transformed model of two tunnel coupled levels attached to two wide-band leads.}
	\label{fig:nrlschem}
\end{figure}

I will now apply the Markovian embedding to the study of two exemplary cases. First, a spinless, noninteracting electronic level attached to two electronic leads $L$ and $R$ (the left and the right) will be considered; it is referred to as the noninteracting resonant level model (see a scheme of the model in Fig.~\ref{fig:nrlschem}). The Hamiltonian of this model reads
\begin{align} \nonumber
\hat{H}_{NRL}=\hat{H}_S+\hat{H}_E+\hat{V}_{SE},
\end{align}
where
\begin{align}
\hat{H}_S&=\epsilon_s C^\dagger C, \\
\hat{H}_E&=\sum_{\alpha=L,R} \sum_k \epsilon_{\alpha k} c^\dagger_{\alpha k} c_{\alpha k}, \\
\hat{V}_{SE}&=\sum_{\alpha=L,R} \sum_k \left( t_{\alpha k} C^\dagger c_{\alpha k} + \text{h.c.} \right).
\end{align}
Here the same notation as in Sec.~\ref{sec:rcm} is applied. The left lead is characterized by a constant spectral density $\Gamma_L$, whereas the right lead is characterized by a Lorentzian spectral density with the amplitude $\gamma_R$ and the bandwidth $2 \delta_R$:
\begin{align}
\Gamma_R(\epsilon)=\frac{\gamma_{R} \delta_R^2}{(\epsilon-\varepsilon_R)^2+\delta_R^2}.
\end{align}
As shown by Zedler et al.~\cite{zedler2009}, such spectral density may lead to a strongly non-Markovian dynamics.
%
%

The Hamiltonian of the model can be transformed using the reaction coordinate mapping to the auxiliary form~\cite{zedler2009, strasberg2016, strasberg2018}
\begin{align} \nonumber
\hat{H}_{NRL}^\text{aux}=\hat{H}_{SA}+\hat{H}_B+\hat{V}_{SA,B}.
\end{align}
Here the Hamiltonian of the supersystem $SA$,
\begin{align} \label{hamsanrl}
\hat{H}_{SA}=\epsilon_s C^\dagger C+\varepsilon_R D^\dagger D+\lambda (C^\dagger D+D^\dagger C),
\end{align}
describes two tunnel-coupled fermionic levels with the coupling integral
\begin{align}
\lambda = \sqrt{\frac{\gamma_R \delta_R}{2}}.
\end{align}
The supersystem is coupled to the residual environment, i.e.,
\begin{align} \label{resbathnrl}
\hat{H}_B&=\sum_k \epsilon_{L k} c^\dagger_{L k} c_{L k}+\sum_k \varepsilon_{X k} d^\dagger_{X k} d_{X k}, \\ \label{rescoupnrl}
\hat{V}_{SA,B}&=\sum_k \left( t_{L k} C^\dagger c_{L k} + \text{h.c.} \right) \\ \nonumber &+\sum_k \left( T_{X k} D^\dagger d_{X k} + \text{h.c.} \right),
\end{align}
where the left lead is unchanged whereas the right lead is replaced by the residual lead $X$ characterized by a constant spectral density
\begin{align}
\Gamma_X=2 \delta_R.
\end{align}

I will now apply the infinite bias limit with $\mu_L \rightarrow \infty$ and $\mu_R \rightarrow -\infty$, where $\mu_\alpha$ is the chemical potential of the lead $\alpha$. Then, as shown by Gurvitz and Prager~\cite{gurvitz1996, gurvitz1998}, the dynamics of the supersystem $SA$ is described by the exact Markovian master equation
\begin{align} \label{gurvprag}
\frac{d}{dt} \rho_{SA} = -\frac{i}{\hbar} \left[\hat{H}_{SA} , \rho_{SA} \right]+\sum_{\alpha=L,X} \mathcal{D}_\alpha \rho_{SA},
\end{align}
where the dissipators $\mathcal{D}_\alpha$ are defined as
\begin{align} \label{disslnrl}
\mathcal{D}_L \rho_{SA} &= \frac{\Gamma_L}{\hbar} \left( C^\dagger \rho_{SA} C -\frac{1}{2} \left\{ C C^\dagger, \rho_{SA} \right\} \right), \\ \label{dissxnrl}
\mathcal{D}_X \rho_{SA} &= \frac{\Gamma_X}{\hbar} \left( D \rho_{SA} D^\dagger - \frac{1}{2} \left\{ D^\dagger D, \rho_{SA} \right\} \right).
\end{align}

The current flowing to the left lead is a dissipative current which corresponds to an incoherent tunneling. The corresponding operator is expressed as
\begin{align}
\hat{I}_L \rho_{SA} = \frac{q \Gamma_L}{\hbar} C^\dagger \rho_{SA} C,
\end{align}
where $q$ is the elementary charge. On the other hand, the current to the right lead is a coherent current, related to the tunneling between the system and the auxiliary site. The corresponding current operator reads
\begin{align}
\hat{I}_R &=-\frac{iq}{\hbar}[\lambda \left(C^\dagger D+ D^\dagger C \right),C^\dagger C] \\ \nonumber &=\frac{iq\lambda}{\hbar} \left(C^\dagger D - D^\dagger C \right).
\end{align}

Matrix representations of the Liouvillian and the current operators are presented in the Appendix~\ref{sec:repr}.

\subsection{Results}
\subsubsection{Dependence on the bandwidth}
%
\begin{figure} 
	\centering
	\includegraphics[width=0.9\linewidth]{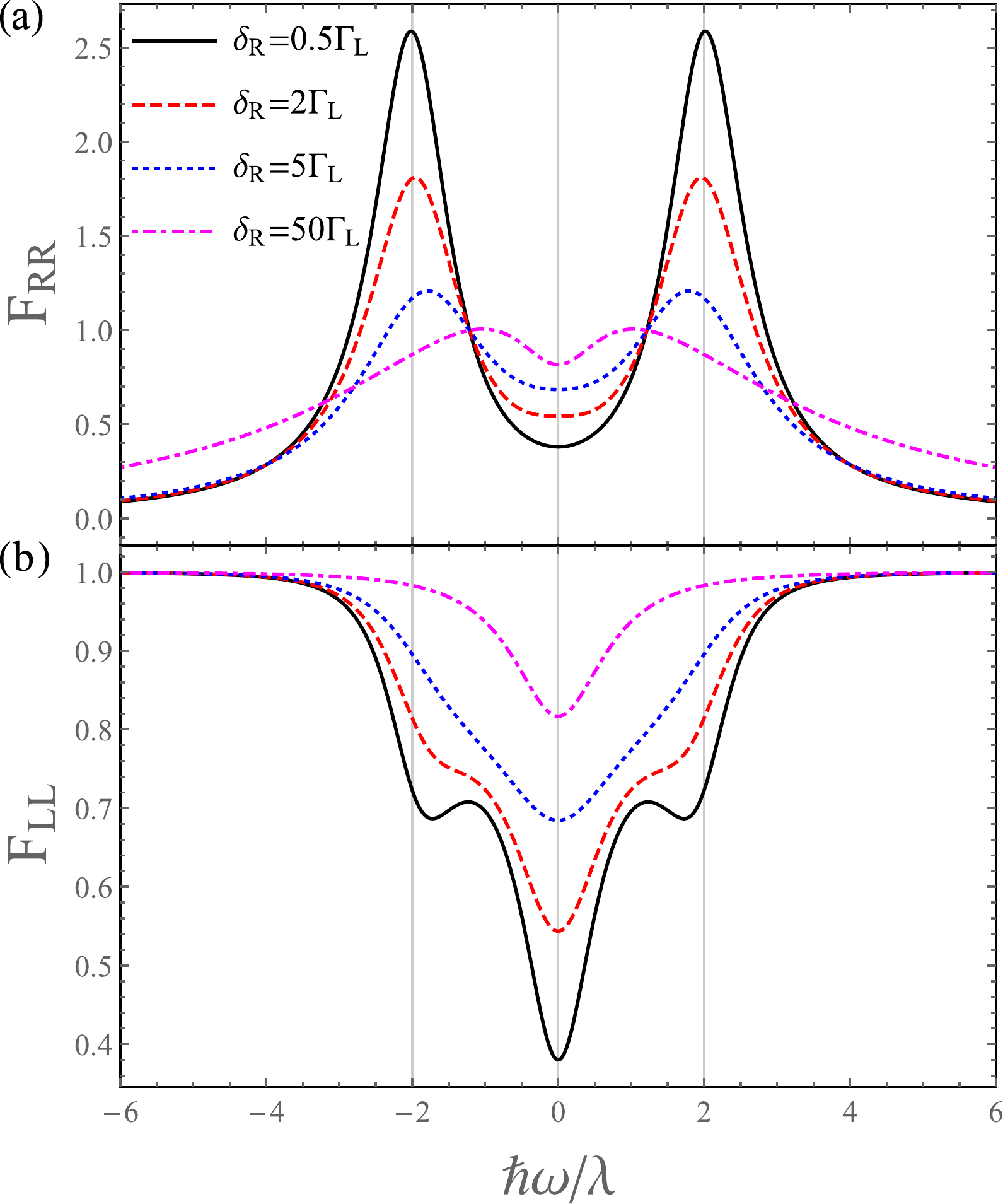} 
	\caption{Frequency-dependent Fano factors $F_{RR}$ (a) and $F_{LL}$ (b) for different values of $\delta_R$ with $\epsilon_s=\varepsilon_R$ and $\gamma_R=10\Gamma_L$.}
	\label{fig:nrlnoise}
\end{figure}
%
I start the analysis with the case when the level energy is in resonance with the center of the Lorentzian band of the right lead ($\epsilon_s=\varepsilon_R$), for which the noise is found to be frequency-symmetric. The autocorrelation noise will be characterized by the (frequency-dependent) Fano factor
\begin{align}
F_{\alpha \alpha} (\omega)=\frac{S_{\alpha \alpha}(\omega)}{|q \langle \hat{I}_\alpha \rangle |},
\end{align}
which is a dimensionless quantity. I will later denote $F_{\alpha \alpha}(\omega)=F_{\alpha \alpha}$ for simplicity. The frequency-dependence of the Fano factors $F_{RR}$ and $F_{LL}$ for different values of $\delta_R$ is presented in Fig.~\ref{fig:nrlnoise}. As one can see in Fig.~\ref{fig:nrlnoise}~(a), for a small $\delta_R$ the Fano factor $F_{RR}$ exhibits pronounced peaks which are located at the frequencies $\omega= \pm 2\lambda/\hbar$; this corresponds to the angular frequency of the coherent oscillations between the levels of the auxiliary system. Similar features for systems undergoing the internal coherent oscillations have been observed in Refs.~\cite{braun2006, luo2007, lambert2008, dong2008, jin2013, droste2015, xue2018, kirton2012, flindt2008, marcos2011}. For larger bandwidths $2 \delta_R$ the peaks diminish and move to lower frequencies, which can be understood as a result of the increased damping of the oscillations by the residual bath $X$; however, they are still visible even for a relatively large bandwidth $2 \delta_R=100 \Gamma_L=10 \gamma_R$. On the other hand, the noise spectrum of the current flowing from the left lead [Fig.~\ref{fig:nrlnoise}~(b)] exhibits some characteristic features such a dips ($\delta_R=0.5 \Gamma_L$) or shoulders ($\delta_R=2 \Gamma_L$) only in the strongly non-Markovian regime when the bandwidth $2 \delta_R$ is smaller than the band amplitude $\gamma_R$. One may also observe that the Fano factor $F_{RR}$ goes to 0 when $\omega \rightarrow \pm \infty$, which is due to the finite width of the energy band; on the other hand, $F_{LL}$ converges to the Poissonian value $F_{LL}=1$, which is characteristic for uncorrelated incoherent tunneling processes.
%
\begin{figure} 
	\centering
	\includegraphics[width=0.9\linewidth]{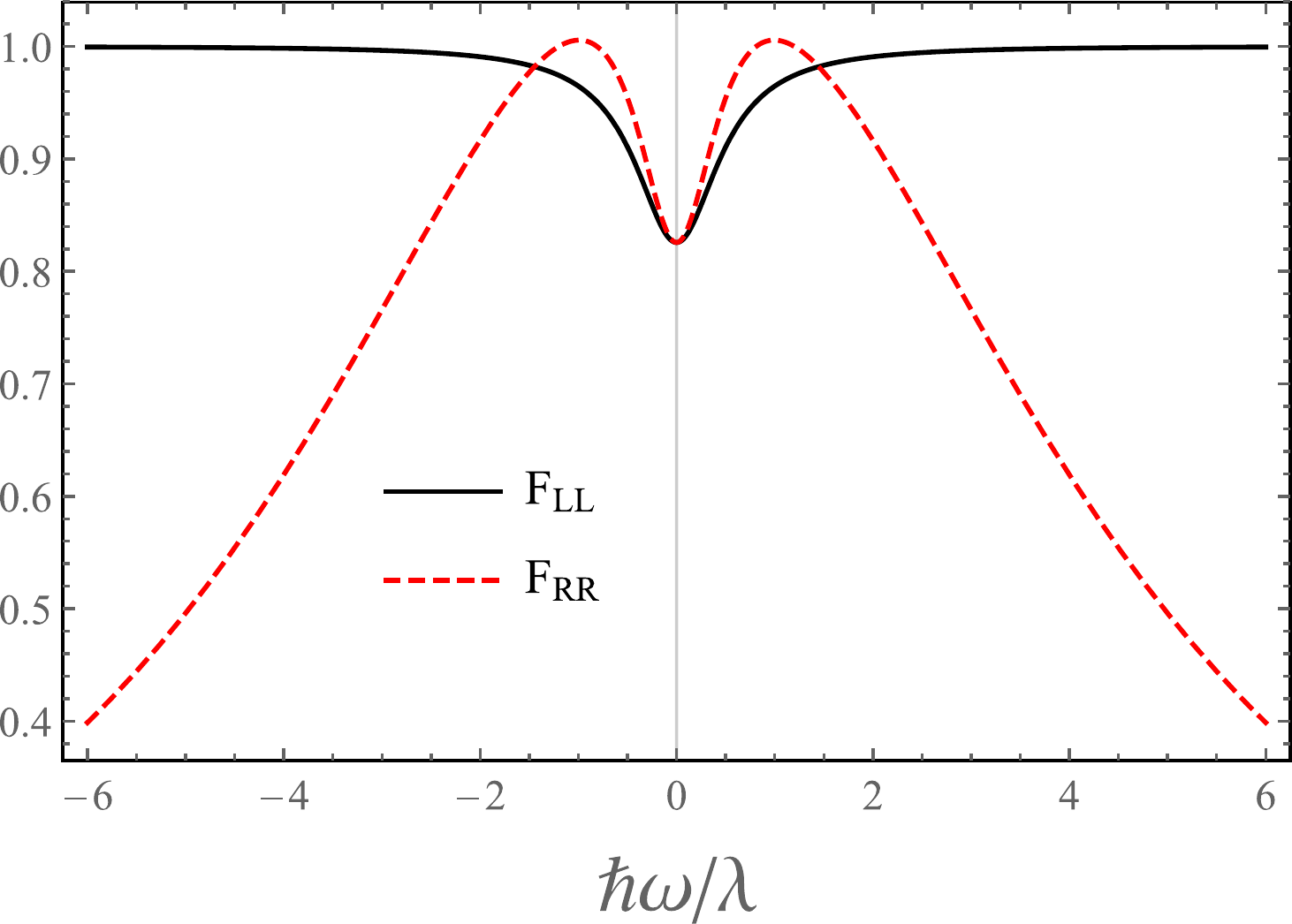} 
	\caption{Frequency-dependent Fano factors $F_{LL}$ and $F_{RR}$ for $\delta_R=50 \Gamma_L$, $\gamma_R=10\Gamma_L$ and $\epsilon_s=\varepsilon_R$.}
	\label{fig:nrlnoisecomp}
\end{figure}
%

Let us now focus for a while on the case when the bandwidth $2\delta_R$ is finite but relatively large with $\gamma_R=10 \Gamma_L$ and $\delta_R=50 \Gamma_{L}$ (Fig.~\ref{fig:nrlnoisecomp}). One may check that in such a case the Fano factor $F_{LL}$ is well approximated by the Markovian expression for a single-level in the wide-band limit ($\delta_R \rightarrow \infty$):
\begin{align} \label{noisemar}
F_{LL}^{\text{Mar}}=F_{RR}^{\text{Mar}}=\frac{\Gamma_L^2+\gamma_R^2+\hbar^2 \omega^2}{\left(\Gamma_L+\gamma_R \right)^2+\hbar^2 \omega^2}.
\end{align}
On the other hand, the Fano factor $F_{RR}$ still deviates from $F_{LL}$, and thus from the Markovian limit. However, both quantities converge for $\omega=0$. This shows that the finite-frequency noise of the current flowing to the energy-dependent right lead is much stronger indicator of the non-Markovianity than the noise of the current flowing from the wide-band left lead and the zero-frequency noise analyzed by Zedler et al.~\cite{zedler2009}.

%
\begin{figure} 
	\centering
	\includegraphics[width=0.9\linewidth]{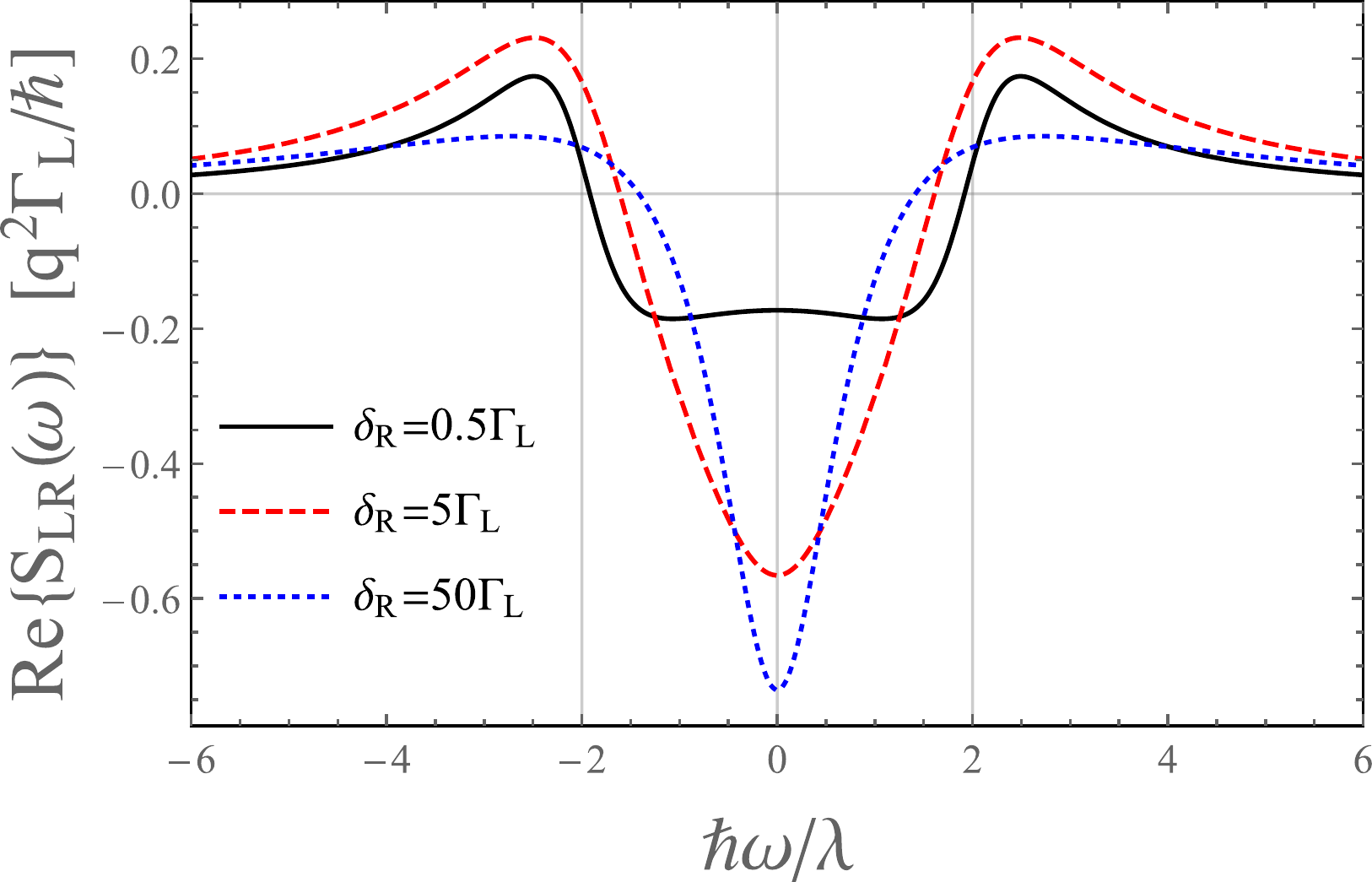} 
	\caption{Real part of the cross-correlation noise $S_{LR}(\omega)$ for different values of $\delta_R$ with $\epsilon_s=\varepsilon_R$ and $\gamma_R=10\Gamma_L$.}
	\label{fig:nrlcrosscor}
\end{figure}
%
The behavior of the real part of the cross-correlation noise is presented in Fig.~\ref{fig:nrlcrosscor} (note that $\text{Re} \{S_{LR}(\omega)\}=\text{Re} \{S_{RL}(\omega)\}$). As one may observe, it is negative for low frequencies (note that the currents are defined as flowing from the system to the lead), whereas for higher frequencies it becomes positive and exhibits peaks in the vicinity of $\omega=\pm 2\lambda/\hbar$. This differs from the Markovian behavior for $\delta_R \rightarrow \infty$ when the real part of the cross-correlation noise is always negative and equal to
\begin{align}
\text{Re} \left\{S_{LR}^\text{Mar}(\omega) \right\}=-\frac{q^2}{\hbar} \frac{\Gamma_L \gamma_R (\Gamma_L^2+\gamma_R^2)}{(\Gamma_L+\gamma_R)^3+(\Gamma_L+\gamma_R) \hbar^2 \omega^2}.
\end{align}
%
\begin{figure} 
	\centering
	\includegraphics[width=0.9\linewidth]{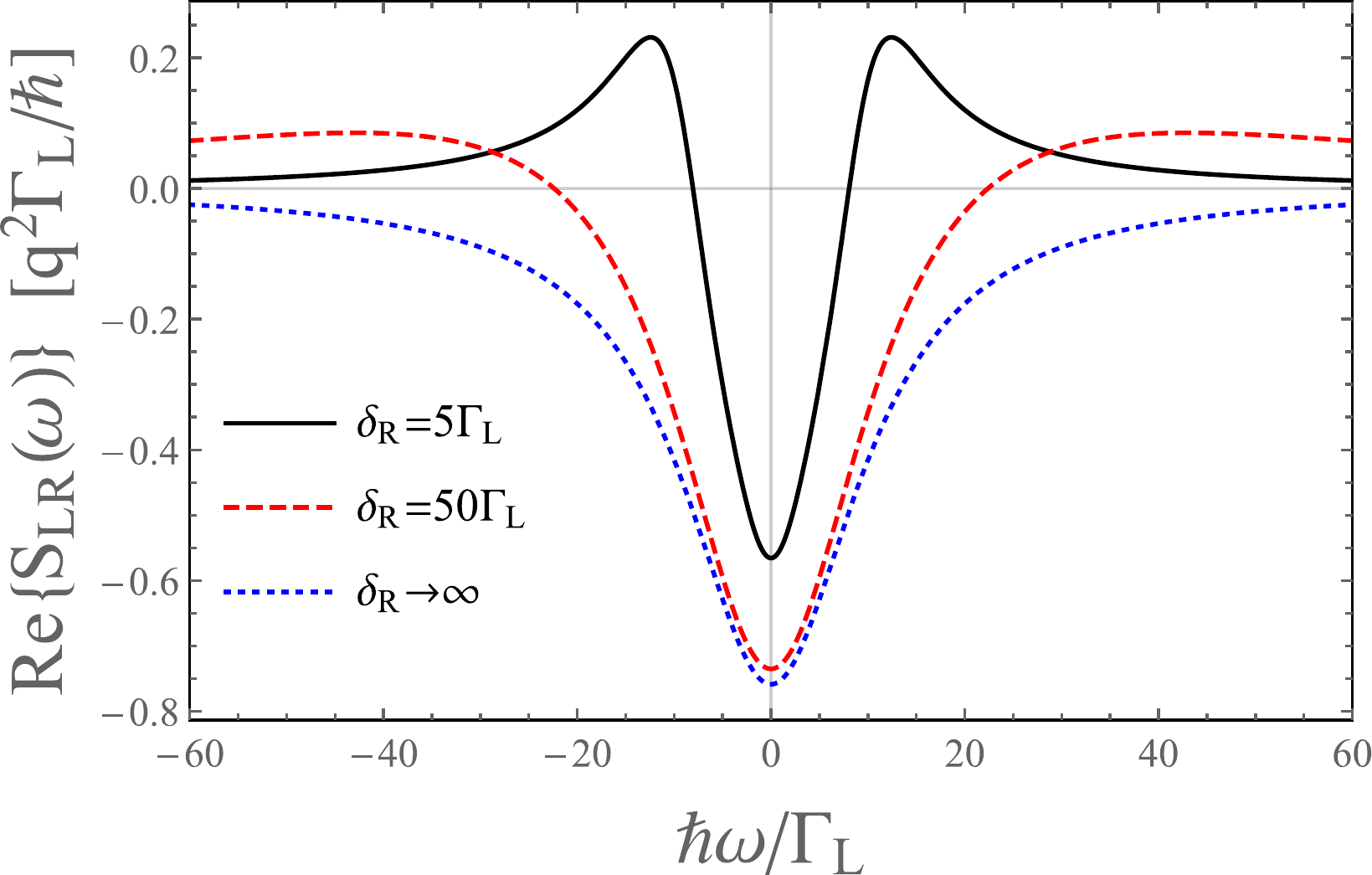} 
	\caption{Real part of the cross-correlation noise $S_{LR}(\omega)$ for different values of $\delta_R$ compared with the Markovian limit $\delta_R \rightarrow \infty$. Parameters as in Fig.~\ref{fig:nrlcrosscor}.}
	\label{fig:nrlcrosscorcomp}
\end{figure}
%
The comparison with the Markovian case is presented in Fig.~\ref{fig:nrlcrosscorcomp} (note a different scale on the $x$-axis). As shown, the cross-correlation noise differs from the Markovian one even for a relatively large bandwidth $2 \delta_R=100 \Gamma_L=10 \gamma_R$. Therefore, its behavior is also a good indicator of non-Markovianity. 

\subsubsection{Noise asymmetry} \label{sec:asym}
%
\begin{figure} 
	\centering
	\includegraphics[width=0.9\linewidth]{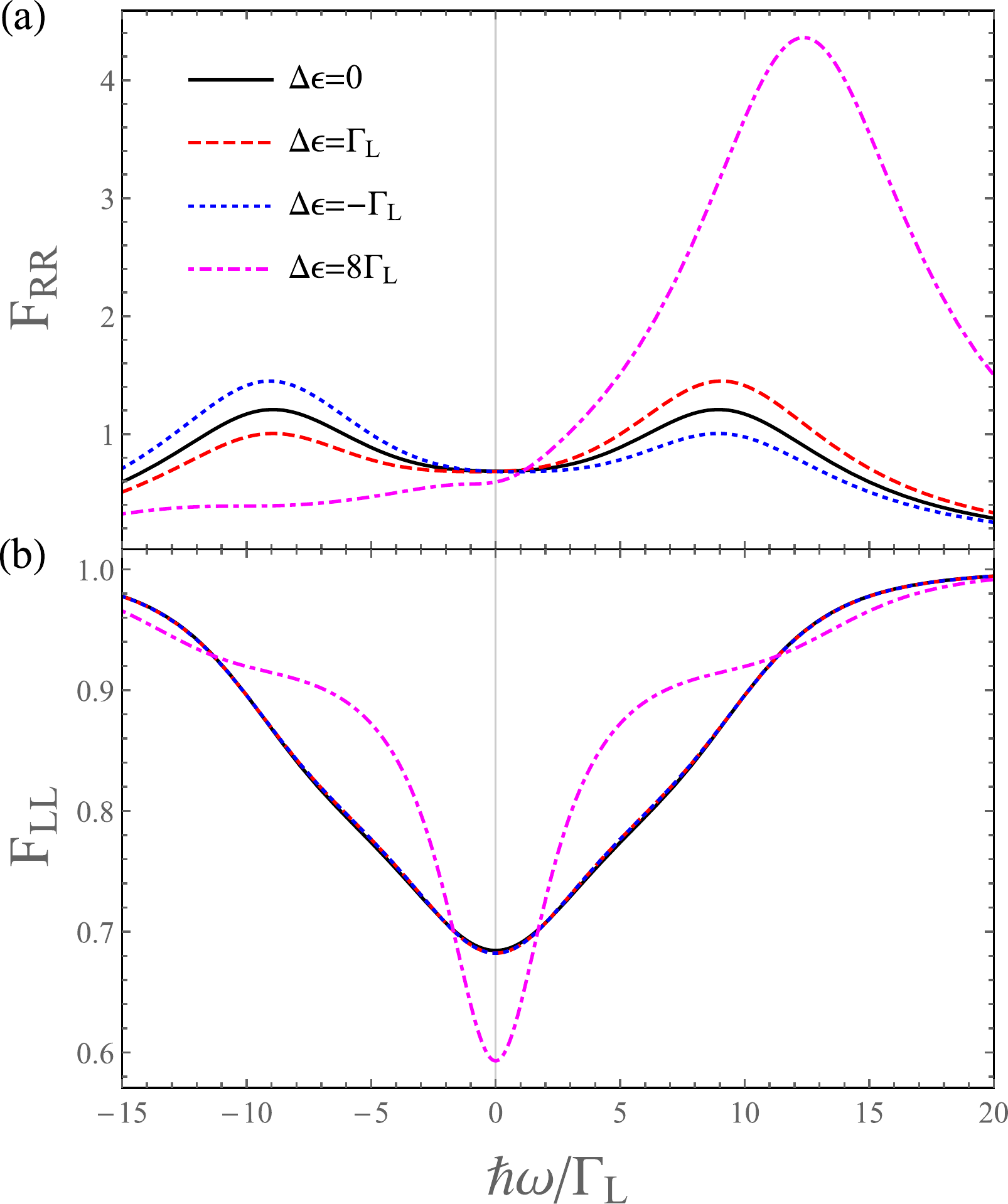} 
	\caption{Frequency-dependent Fano factors $F_{RR}$ (a) and $F_{LL}$ (b) for different values of $\Delta \epsilon=\epsilon_s-\varepsilon_R$ with $\gamma_R=10\Gamma_L$ and $\delta_R=5\Gamma_L$.}
	\label{fig:nrlnoiseeps}
\end{figure}
%
The noise spectra presented so far have been symmetric with respect to the frequency, which is a consequence of the resonance between the system level and the center of the Lorentzian band. As shown in Fig.~\ref{fig:nrlnoiseeps}, the Fano factor $F_{RR}$ becomes asymmetric when the level is shifted from the resonance; here the energy shift is parametrized by $\Delta \epsilon=\epsilon_s-\varepsilon_R$ and a moderately large bandwidth $2 \delta_R=\gamma_R=10 \Gamma_L$ is taken. Specifically, for $\Delta \epsilon>0$ ($\Delta \epsilon<0$) the emission (absorption) band is enhanced. This is a result of the inelastic nature of the tunneling: for $\Delta \epsilon>0$ and $\Delta \epsilon<0$ the electrons tunneling to the right lead lose some energy or absorb some energy, respectively. As one may observe, the noise is significantly affected even for a relatively small energy shift $\Delta \epsilon= \pm \Gamma_L$. For a larger shift $\Delta \epsilon =8 \Gamma_L$ the emission peak becomes even more pronounced and shifts to a higher frequency $\hbar \omega \approx \sqrt{\Delta \epsilon^2+(2\lambda)^2}$, which corresponds to the energy splitting of the singly-occupied eigenstates of the auxiliary double-level model; cf. a similar result in Ref.~\cite{flindt2008}.

The Fano factor $F_{LL}$, on the other hand, is nearly not affected for a small energy shift $\Delta \epsilon= \pm \Gamma_L$. For a larger shift one may observe that the zero-frequency dip becomes deeper and narrower. The suppression of the zero-frequency noise occurs because by moving the energy level out of the band center one reduces the tunneling rate to the right lead. This increases the symmetry of the tunnel couplings to the left and the right lead, leading to the noise suppression [cf. Eq.~\eqref{noisemar} for the Markovian case]. Accordingly, the narrowing of the dip is a result of the reduced dissipation rate~\cite{bulka1999}.

%
\begin{figure} 
	\centering
	\includegraphics[width=0.9\linewidth]{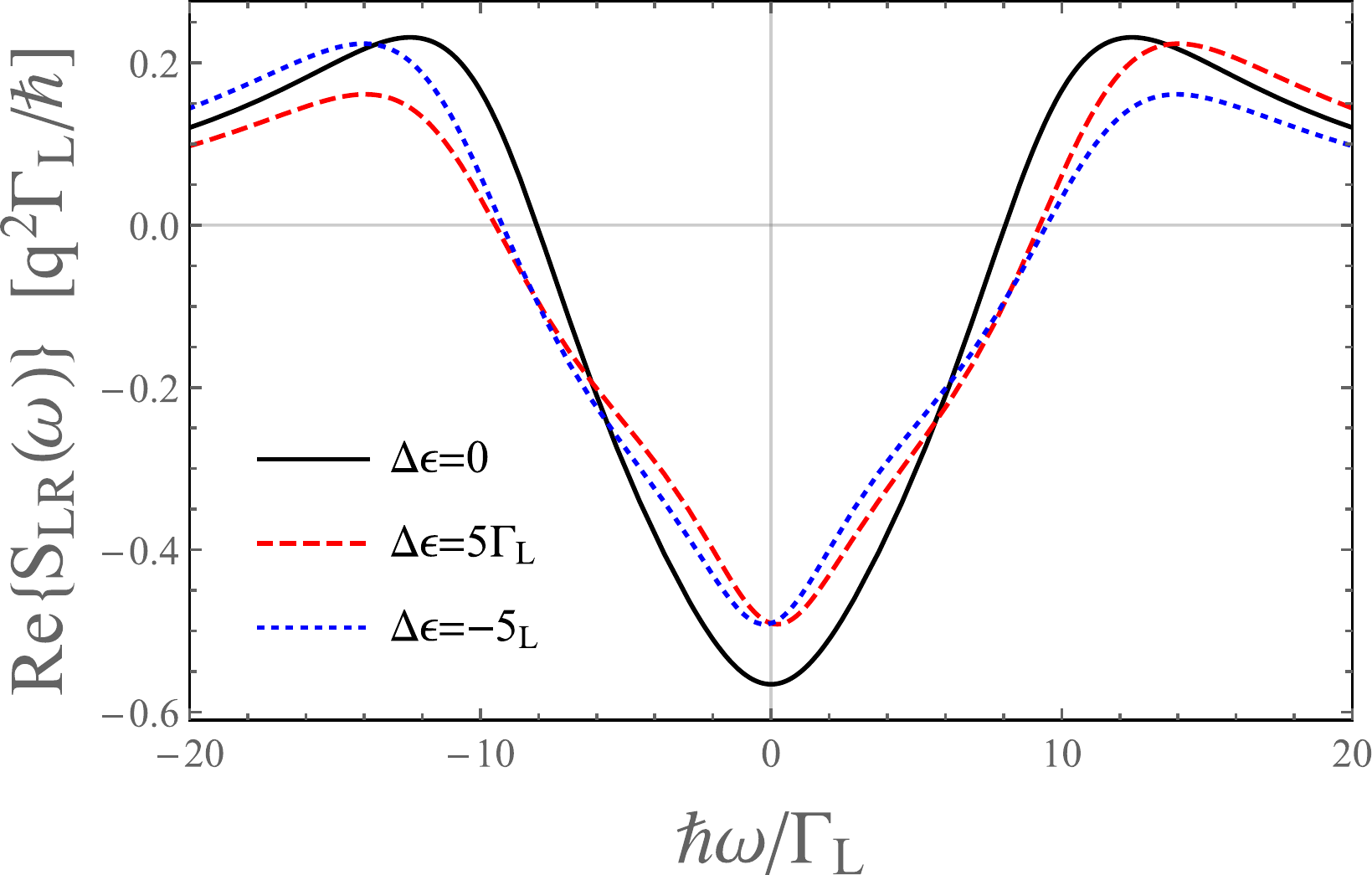} 
	\caption{Real part of the cross-correlation noise $S_{LR}(\omega)$ for different values of $\Delta \epsilon=\epsilon_s-\varepsilon_R$ with $\gamma_R=10\Gamma_L$ and $\delta_R=5\Gamma_L$.}
	\label{fig:nrlcrosscoreps}
\end{figure}
%
For completeness, the behavior of the cross-correlation noise is presented in Fig.~\ref{fig:nrlcrosscoreps}. As one may check, it is nearly not affected for a small shift $\Delta \epsilon= \pm \Gamma_L$ (not shown). For a larger shift $\Delta \epsilon =\pm 5 \Gamma_L$ one may observe the noise asymmetry, which is, however, not as pronounced as in the case of $F_{RR}$. 

\subsubsection{Agreement with the Green's function results} \label{sec:compgreen}
Finally, let me mention that the presented results have been fully reproduced using the nonequilbrium Green's function formalism. Specifically, to calculate the noise the formulas derived by Zamoum et al.~\cite{zamoum2016} have been applied; this approach is presented in the Appendix~\ref{sec:negf}. Since the agreement is complete, there is no need to discuss the results in detail.

\section{Anderson model} \label{sec:and}
\subsection{Model}
So far I have analyzed a noninteracting system which can be described exactly using the Green's function formalism (also for finite voltages)~\cite{zamoum2016}. To demonstrate the full power of the proposed approach, let us now consider a paradigmatic example of a strongly interacting system, namely, the Anderson model. It is described by the Hamiltonian
\begin{align} \nonumber
\hat{H}_\text{And}=\hat{H}_S+\hat{H}_E+\hat{V}_{SE},
\end{align}
where the system Hamiltonian
\begin{align}
\hat{H}_S =\epsilon_s \sum_{\sigma} C_\sigma^\dagger C_\sigma+ U C^\dagger_\uparrow C_\uparrow C^\dagger_\downarrow C_\downarrow
\end{align}
describes a single electronic site with an on-site Coulomb interaction $U$; here $C_\sigma^\dagger$ is the creation operator of an electron with a spin $\sigma \in \{\uparrow, \downarrow\}$. Analogously, the bath and the tunneling Hamiltonians~\eqref{resbathnrl} and~\eqref{rescoupnrl} are replaced by their spin-dependent counterparts
\begin{align}
\hat{H}_E&=\sum_{\alpha=L,R} \sum_{k \sigma} \epsilon_{\alpha k \sigma} c^\dagger_{\alpha k \sigma} c_{\alpha k \sigma}, \\
\hat{V}_{SE}&=\sum_{\alpha=L,R} \sum_{k \sigma} \left( t_{\alpha k} C^\dagger_\sigma c_{\alpha k \sigma} + \text{h.c.} \right),
\end{align}
while the coupling to the lead is taken to be spin-independent:
\begin{align}
\Gamma_L^\sigma(\epsilon) &= 2 \pi \sum_k |t_{\alpha k} |^2 \delta (\epsilon-\epsilon_{Lk\sigma})= \Gamma_L, \\
\Gamma_R^\sigma (\epsilon) &=2 \pi \sum_k |t_{\alpha k} |^2 \delta (\epsilon-\epsilon_{Rk\sigma})=\frac{\gamma_{R} \delta_R^2}{(\epsilon-\varepsilon_R)^2+\delta_R^2}.
\end{align}

Since the procedure used to transform the Hamiltonian is the same as in the previous case, it is enough to present the final results. The Hamiltonian of the supersystem $SA$ reads now
\begin{align}
\hat{H}_{SA}&=\epsilon_s \sum_{\sigma} C_{\sigma}^\dagger C_\sigma+ U C^\dagger_\uparrow C_\uparrow C^\dagger_\downarrow C_\downarrow \\ \nonumber &+\varepsilon_R \sum_{\sigma} D^\dagger_\sigma D_\sigma+\lambda \sum_\sigma (C_\sigma^\dagger D_\sigma+D_\sigma^\dagger C_\sigma),
\end{align}
with $\lambda = \sqrt{\gamma_R \delta_R/2}$ (same as before); it thus describes the interacting system tunnel-coupled to the noninteracting auxiliary site. The dynamics is described by Eq.~\eqref{gurvprag} with spinfull dissipators
\begin{align}
\mathcal{D}_L \rho_{SA} &= \frac{\Gamma_L}{\hbar} \sum_{\sigma } \left( C_\sigma^\dagger \rho_{SA} C_\sigma -\frac{1}{2} \left\{ C_\sigma C_\sigma^\dagger, \rho_{SA} \right\} \right), \\
\mathcal{D}_X \rho_{SA} &= \frac{\Gamma_X}{\hbar} \sum_{\sigma} \left( D_\sigma \rho_{SA} D_\sigma^\dagger - \frac{1}{2} \left\{ D_\sigma^\dagger D_\sigma, \rho_{SA} \right\} \right),
\end{align}
where $\Gamma_X=2 \delta_R$, and the current operators are defined as
\begin{align}
\hat{I}_L \rho_{SA} &= \frac{q \Gamma_L}{\hbar} \sum_{\sigma} C_\sigma^\dagger \rho_{SA} C_\sigma, \\
\hat{I}_R &=\frac{iq \lambda}{\hbar} \sum_{\sigma} \left(C_\sigma^\dagger D_\sigma - D_\sigma^\dagger C_\sigma \right).
\end{align}
The matrix representations of the Liouvillian and the current operators are now too large to be presented explicitly in the paper. However, the procedure to obtain them is briefly sketched in the Appendix~\ref{sec:repr}.

\subsection{Results}
%
\begin{figure} 
	\centering
	\includegraphics[width=0.9\linewidth]{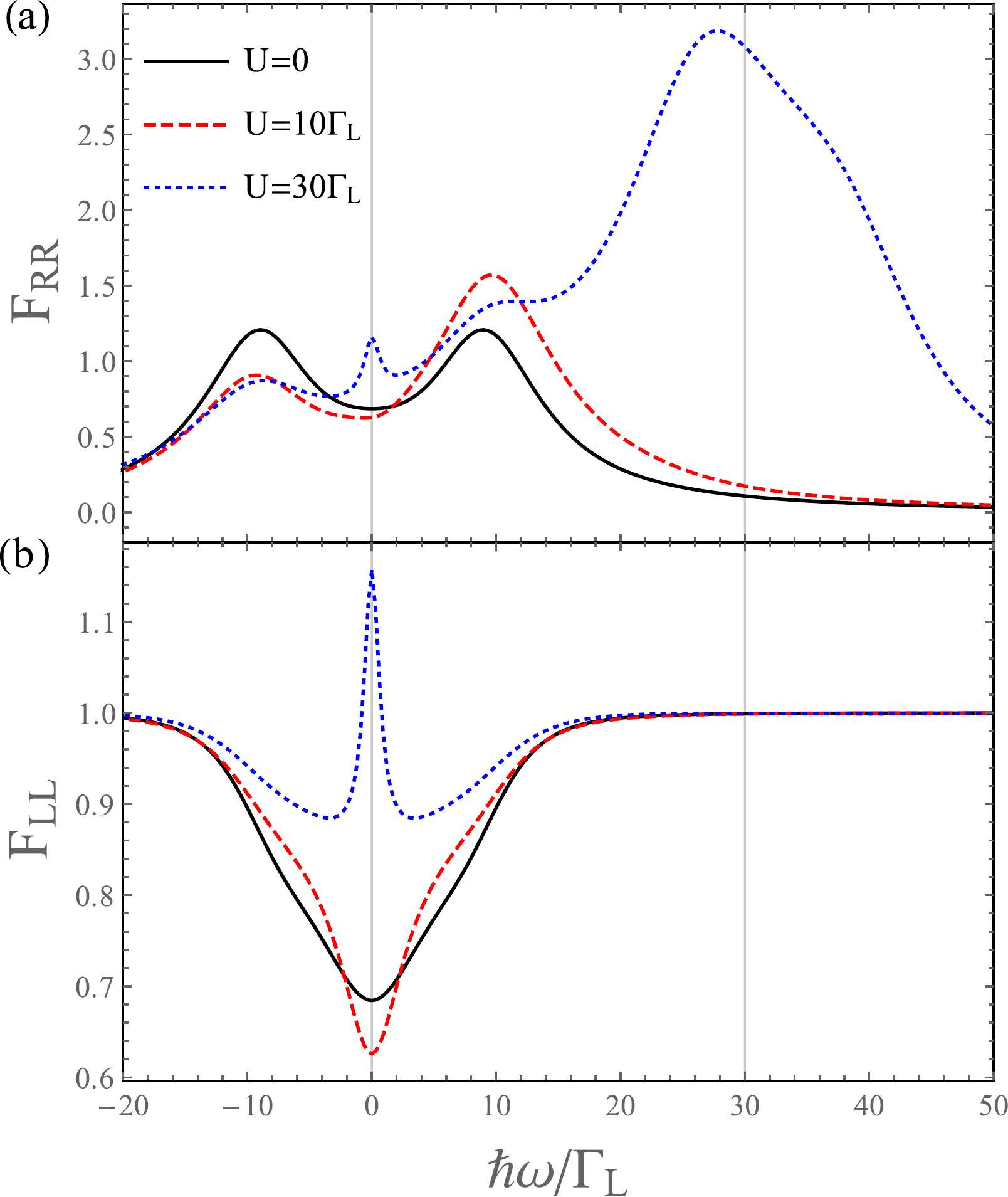} 
	\caption{Frequency-dependent Fano factors $F_{RR}$ (a) and $F_{LL}$ (b) for different values of $U$ with $\epsilon_s=\varepsilon_R$, $\gamma_R=10\Gamma_L$ and $\delta_R=5\Gamma_L$.}
	\label{fig:andnoise}
\end{figure}
%
The frequency-dependence of the Fano factors $F_{RR}$ and $F_{LL}$ for different values of the Coulomb coupling $U$ is presented in Fig.~\ref{fig:andnoise}. Specifically, the case of a moderately large bandwidth $2 \delta_R=\gamma_R=10 \Gamma_L$ is analyzed. As one may observe, the Coulomb coupling leads to the asymmetry of $F_{RR}$ by enhancing the emission peak [Fig.~\ref{fig:andnoise}~(a)]. This is because the Coulomb repulsion increases the energy of the electrons tunneling to the right lead from a doubly-occupied system. For large $U=30 \Gamma_L$ the emission peak is very pronounced and shifted to the frequency $\hbar \omega \approx U$. This indicates that for a large Coulomb coupling the value of $U$ rather than $\lambda = 5 \Gamma_L$ determines the position of the emission peak.

The behavior of $F_{LL}$ [Fig.~\ref{fig:andnoise}~(b)] is quite non-trivial: the zero-frequency noise becomes suppressed for a moderate $U=10 \Gamma_L$, whereas for a large $U=30 \Gamma_L$ it is enhanced to a super-Poissonian value $F_{LL} > 1$. The former case can be explained by the same mechanism as discussed in Sec.~\ref{sec:asym}: the Coulomb repulsion shifts the on-site electron energy from the center of the Lorentzian band, thus reducing the effective tunneling rate to the right lead and the coupling asymmetry. Apparently, the energy-dependence of the tunneling rate is also responsible for the noise enhancement for $U=30 \Gamma_L$. This is because the system switches between the singly-occupied state, for which the tunneling rate to the right lead is high, and the doubly-occupied state, for which the tunneling is suppressed. Such a switching between the states with a high and a low conductance, referred to as the dynamical channel blockade, leads to the noise enhancement~\cite{bulka2000, belzig2005}.

%
\begin{figure} 
	\centering
	\includegraphics[width=0.9\linewidth]{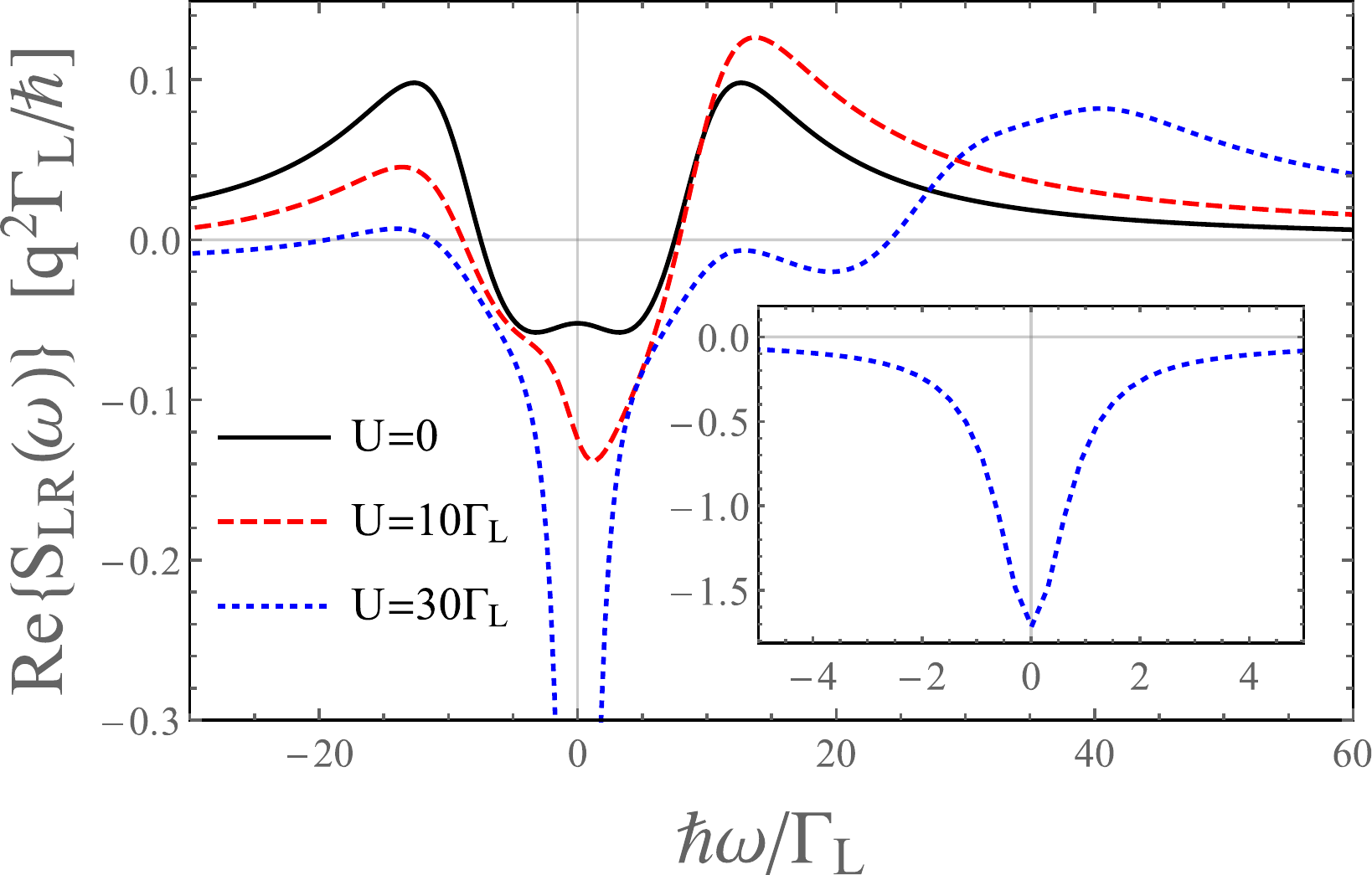} 
	\caption{Real part of the cross-correlation noise $S_{LR}(\omega)$ for different values of $U$ with $\epsilon_s=\varepsilon_R$, $\gamma_R=10\Gamma_L$ and $\delta_R=5\Gamma_L$. The smaller panel presents a zero-frequency dip for $U=30 \Gamma_L$.}
	\label{fig:andcrosscor}
\end{figure}
%
As one may expect, the Coulomb interaction affects also the cross-correlation noise leading to its asymmetry (Fig.~\ref{fig:andcrosscor}). The most pronounced effects are observed for a large coupling $U=30 \Gamma_L$. First, as for $F_{RR}$, the peak of the noise spectrum moves to a higher frequency. Secondly, one may observe a pronounced dip for $\omega=0$, which is fully presented in the smaller panel. The latter effect seems to be associated with the aforementioned dynamical channel blockade mechanism.

\section{Conclusions} \label{sec:concl}
The paper has presented the approach to calculate the finite-frequency quantum noise in non-Markovian systems based on the Markovian embedding. Its applicability has been demonstrated using the model of a single electronic level coupled to the bath with a strongly energy-dependent density of states which leads to a non-Markovian behavior; this system has been mapped onto an equivalent, exactly-solvable double-level model. The applied method has been shown to reveal genuinely quantum effects, such as the noise asymmetry caused by the inelastic nature of the tunneling to the energy-dependent lead. Furthermore, it is applicable also in the strongly non-Markovian regime for which the approximate master equation approaches have been shown to fail~\cite{zedler2009}. Finally, the technique enables one to study the role of the Coulomb interaction and the non-Markovian dynamics on equal footing, revealing the phenomena characteristic for interacting systems, such as the super-Poissonian noise enhancement due to the dynamical channel blockade. This contrasts with most common Green's function or master equation approaches which treat either the intra-system interactions or the coupling to the bath in an approximate way.

It should be emphasized that the reaction coordinate mapping applied in this paper provides a mapping on the exactly-solvable double level model only for a Lorentzian spectral density of the bath; therefore, this approach may be not easily applicable to realistic systems with more complex spectral densities. However, the case analyzed may be used as a benchmark for approximate methods. Furthermore, the presented procedure to calculate the noise may be implemented using more advanced Markovian embedding techniques~\cite{chen2019, chen2019b}, such as the auxiliary master equation which has been applied for the study of correlated impurities in the Kondo regime~\cite{dorda2014, arrigoni2013}. Further generalizations of the presented approach may also include the calculation of other types of noise, such as heat or mixed charge-heat noises~\cite{crepieux2015}.

\begin{acknowledgments}
I thank B. R. Bu\l{}ka for the valuable discussion and comments on the manuscript. The author has been supported by the National Science Centre, Poland, under the project No.~2017/27/N/ST3/01604, by the the START scholarship of the Foundation for Polish Science (FNP) and by the Scholarships of Minister of Science and Higher Education.
\end{acknowledgments}

\appendix

\section{Matrix representation of the noise formula} \label{sec:repr}
\subsection{Hilbert space representation}
In this section a practical procedure to calculate the noise using Eq.~\eqref{noiseform} is presented. To this goal one first needs to find a representation of the Liouvillian in the Hilbert space where the density matrix is represented by a square matrix. Here I focus on the case when both the Hamiltonian and the dissipators can be expressed in terms of creation and annihilation operators [see, e.g., Eqs.~\eqref{hamsanrl}, \eqref{disslnrl} and~\eqref{dissxnrl}]. For a system with $N$ fermionic sites $i$ these operators can be conveniently represented using the Jordan-Wigner transformation~\cite{schaller2014}:
\begin{align}
c_i^\dagger&=\bigotimes_{k=1}^{i-1} \sigma^z \otimes \sigma^+ \otimes \bigotimes_{k=i+1}^N \mathds{1}_2, \\
c_i&=\bigotimes_{k=1}^{i-1} \sigma^z \otimes \sigma^- \otimes \bigotimes_{k=i+1}^N \mathds{1}_2,
\end{align}
where
\begin{align}
\mathds{1}_2 = & \begin{pmatrix} 1 & 0 \\ 0 & 1 \end{pmatrix}, \\
\sigma^z = & \begin{pmatrix} 1 & 0 \\ 0 & -1 \end{pmatrix}, \\
\sigma^+ = & \begin{pmatrix} 0 & 1 \\ 0 & 0 \end{pmatrix}, \\
\sigma^- = & \begin{pmatrix} 0 & 0 \\ 1 & 0 \end{pmatrix}.
\end{align}

\subsection{Liouville space representation}
It is now convenient to move to the Liouville space in which the density matrix is represented by a column vector $\tilde{\rho}$, whereas the other operators and superoperators $\hat{O}$ are represented by square matrices $\tilde{\hat{O}}$ (with tilde)~\cite{brandes2008}; then, using Eq.~\eqref{noiseform}, the noise is given by a product of matrices. The matrix $\tilde{\hat{O}}$ can be determined using the equivalence of the Hilbert space and the Liouville space representations:
\begin{align}
\text{Tr} \left[\hat{O} \rho \right] = \tilde{\zeta} \tilde{\hat{O}} \tilde{\rho},
\end{align}
where $\tilde{\zeta}$ is a row vector with the elements equal to 1 on positions corresponding to the positions of the diagonal elements of the density matrix in $\tilde{\rho}$ while 0 otherwise. Technically, the matrix $\tilde{\hat{O}}$ can be either found ``by hand'' or by using an appropriate computer algebra system.

More specifically, the vector $\tilde{\rho}$ needs to contain only those elements of the density matrix which are relevant for the dynamics of the system, since some off-diagonal elements may be decoupled from the dynamics of the other, and thus exponentially decay. Let us now focus on a specific case of the noninteracting resonant level models analyzed in Sec.~\ref{sec:nrl}. The Hilbert space of the system contains four states: the empty state $|0 \rangle$, the state $|S \rangle$ ($|A \rangle$) with the occupied system level (auxiliary level) and the doubly-occupied state $|2 \rangle$. The row vector $\tilde{\rho}$ includes then six relevant elements:
\begin{align}
\tilde{\rho}= \left( \rho_{00}, \rho_{SS}, \rho_{AA}, \rho_{22}, \rho_{SA}, \rho_{AS} \right)^T.
\end{align}
The corresponding vector $\tilde{\zeta}$ reads
\begin{align}
\tilde{\zeta}= \left(1,1,1,1,0,0 \right).
\end{align}
The matrix representation of the Liouvillian takes the form
\begin{align}
\tilde{\mathcal{L}}=\frac{1}{\hbar}
\begin{pmatrix} -\Gamma_L & 0 & \Gamma_X & 0 & 0 & 0 \\ 
\Gamma_L & 0 & 0 & \Gamma_X & i \lambda & -i \lambda \\
0 & 0  & -\Gamma_L-\Gamma_X & 0 & -i \lambda & i \lambda \\
0 & 0 & \Gamma_L & -\Gamma_X & 0 & 0 \\
0 & i \lambda & -i \lambda & 0 &z_- & 0 \\
0 & -i \lambda & i \lambda & 0 & 0 & z_+\end{pmatrix},
\end{align}
where $z_\pm=-(\Gamma_L+\Gamma_X)/2 \pm i (\epsilon_s-\varepsilon_R)$, and the current operators are represented as
\begin{align}
\tilde{\hat{I}}_L&= \frac{q \Gamma_L}{\hbar} \begin{pmatrix}
0 & 0 & 0 & 0 & 0 & 0 \\
1 & 0 & 0 & 0 & 0 & 0 \\
0 & 0 & 0 & 0 & 0 & 0 \\
0 & 0 & 1 & 0 & 0 & 0 \\
0 & 0 & 0 & 0 & 0 & 0 \\
0 & 0 & 0 & 0 & 0 & 0
\end{pmatrix}, \\
\tilde{\hat{I}}_R&= \frac{i q \lambda}{\hbar} \begin{pmatrix}
0 & 0 & 0 & 0 & 0 & 0 \\
0 & 0 & 0 & 0 & 0 & -1 \\
0 & 0 & 0 & 0 & 1 & 0 \\
0 & 0 & 0 & 0 & 0 & 0 \\
0 & 0 & -1 & 0 & 0 & 0 \\
0 & 1 & 0 & 0 & 0 & 0
\end{pmatrix},
\end{align}

\section{Nonequilibrium Green's function noise formula} \label{sec:negf}
As mentioned in Sec.~\ref{sec:compgreen} the results for the noninteracting resonant level model have been fully reproduced using the nonequilibrium Green's function formalism, and more specifically, applying the exact formula for the noise derived by Zamoum et al.~\cite{zamoum2016}. To make the paper self-contained this formula is here rewritten. I will first present different components of the formula. The retarded/advanced Green's function of the noninteracting resonant level reads
\begin{align}
G^{r/a}(\epsilon)&=\frac{1}{\epsilon-\epsilon_s-\Sigma^{r/a}(\epsilon)},
\end{align}
where 
\begin{align}
\Sigma^{r/a}(\epsilon)=\Sigma_L^{r/a}(\epsilon)+\Sigma_R^{r/a}(\epsilon)
\end{align}
is the retarded/advanced self-energy with 
\begin{align}
\Sigma^{r/a}_\alpha (\epsilon)&=\Lambda_\alpha (\epsilon) \mp i \Gamma_\alpha (\epsilon)/2, \\
\end{align}
being the self-energy of the bath $\alpha$; here $\Lambda_\alpha(\epsilon)$ is the level-shift function. For the system considered in the paper $\Lambda_L(\epsilon)=0$ whereas~\cite{zedler2009}
\begin{align}
\Lambda_R(\epsilon)&=\frac{\gamma_R \delta_R}{2} \frac{\epsilon-\varepsilon_R}{(\epsilon-\varepsilon_R)^2+\delta_R^2}.
\end{align}
The lesser/greater Green's function is expressed as
\begin{align}
G^{</>}(\epsilon)&=\pm iA(\epsilon)\tilde{f}^{e/h}(\epsilon), \end{align}
where
\begin{align}
A(\epsilon)&=\frac{\Gamma(\epsilon)}{[\epsilon-\epsilon_s-\Lambda(\epsilon)]^2+[\Gamma(\epsilon)/2]^2},
\end{align}
with
\begin{align}
\Gamma(\epsilon)=&\Gamma_L(\epsilon)+\Gamma_R(\epsilon), \\
\Lambda(\epsilon)=&\Lambda_L(\epsilon)+\Lambda_R(\epsilon),
\end{align}
is the spectral function, whereas
\begin{align}
\tilde{f}^{e/h}_\alpha(\epsilon)&=\frac{\sum_{\alpha} \Gamma_\alpha(\epsilon) f^{e/h}_\alpha(\epsilon)}{\Gamma (\epsilon)}
\end{align}
is the averaged Fermi distribution; here
\begin{align}
f_\alpha^e(\epsilon)=\frac{1}{1+\exp[(\epsilon-\mu_\alpha)/k_B T_\alpha]}
\end{align}
is the Fermi distribution of the lead $\alpha$, where $T_\alpha$ and $\mu_\alpha$ are the corresponding temperature and the chemical potential, and $f^h_\alpha(\epsilon)=1-f_\alpha^e(\epsilon)$.

Using the expressions above the averaged current can be calculated as
\begin{align}
&\langle \hat{I}_L \rangle=-\langle \hat{I}_R \rangle \\ \nonumber &=\frac{q}{2 \pi \hbar} \int_{-\infty}^{\infty} d \epsilon \frac{\Gamma_L (\epsilon) \Gamma_R (\epsilon)}{[\epsilon-\epsilon_s-\Lambda(\epsilon)]^2+[\Gamma(\epsilon)/2]^2} [f_L^e (\epsilon)-f_R^e (\epsilon)].
\end{align}

Finally, the noise is a sum of five contributions,

\begin{align}
S_{\alpha \beta} (\omega) = \sum_{k=1}^5 C^{(k)}_{\alpha \beta}(\omega),
\end{align}
where
\begin{widetext}
\begin{align}
C^{(1)}_{\alpha \beta} (\omega) =\frac{i q^2}{2 \pi \hbar} \delta_{\alpha \beta}  \int_{-\infty}^{\infty} d \epsilon \Gamma_\alpha (\epsilon) [f_\alpha^e(\epsilon) G^>(\epsilon-\hbar \omega)-f_\alpha^h(\epsilon) G^<(\epsilon +\hbar \omega)],
\end{align}

\begin{align}\nonumber &C^{(2)}_{\alpha \beta} (\omega)= -\frac{i q^2}{2 \pi \hbar} \int_{-\infty}^{\infty} d\epsilon \left[ \Gamma_\alpha(\epsilon) G^r (\epsilon) f_\alpha^e (\epsilon) G^> (\epsilon-\hbar \omega) \Sigma^a_\beta (\epsilon-\hbar \omega)- G^< (\epsilon+\hbar \omega) \Sigma_\alpha^a(\epsilon+\hbar \omega) G^r (\epsilon) \Gamma_\beta (\epsilon) f^h_\beta(\epsilon) \right] \\ &-\frac{q^2}{2 \pi \hbar}  \int_{-\infty}^{\infty} d \epsilon  \left[ \Gamma_\alpha (\epsilon) G^r(\epsilon) f_\alpha^e (\epsilon) \Gamma_\beta (\epsilon-\hbar \omega) G^r(\epsilon-\hbar \omega) f_\beta^h (\epsilon-\hbar \omega)+ G^< (\epsilon) \Sigma_\alpha^a(\epsilon) G^> (\epsilon-\hbar \omega) \Sigma_\beta^a(\epsilon-\hbar \omega) \right],
\end{align}

\begin{align}
C^{(3)}_{\alpha \beta} (\omega) =& \frac{i q^2}{2 \pi \hbar} \int_{-\infty}^{\infty} d\epsilon G^> (\epsilon-\hbar \omega) [\Gamma_\alpha(\epsilon) f_\alpha^e(\epsilon) G^r(\epsilon) \Sigma_\beta^r(\epsilon) +\Sigma_\alpha^a(\epsilon) G^a(\epsilon) \Gamma_\beta(\epsilon) f_\beta^e(\epsilon)] \\ \nonumber
&+\frac{q^2}{2 \pi \hbar} \int_{-\infty}^{\infty} d\epsilon G^> (\epsilon-\hbar \omega) G^<(\epsilon) \Sigma_\alpha^a(\epsilon) \Sigma_\beta^r(\epsilon),
\end{align}

\begin{align}
C^{(4)}_{\alpha \beta} (\omega) =&-\frac{i q^2}{2 \pi \hbar} \int_{-\infty}^{\infty} d\epsilon G^< (\epsilon+\hbar \omega) [\Gamma_\alpha(\epsilon) f^h_\alpha (\epsilon) G^a (\epsilon) \Sigma_\beta^a (\epsilon) +\Sigma_\alpha^r (\epsilon) G^r (\epsilon) \Gamma_\beta(\epsilon) f^h_\beta (\epsilon)] \\ \nonumber
&+\frac{q^2}{2 \pi \hbar} \int_{-\infty}^{\infty} d\epsilon G^< (\epsilon+\hbar \omega) G^>(\epsilon) \Sigma_\alpha^r(\epsilon) \Sigma_\beta^a(\epsilon),
\end{align}

\begin{align}\nonumber &C^{(5)}_{\alpha \beta} (\omega)=\frac{i q^2}{2 \pi \hbar}  \int_{-\infty}^{\infty} d \epsilon \left[ G^a (\epsilon) \Gamma_{\alpha}(\epsilon) f^h_\alpha (\epsilon) G^<(\epsilon+\hbar \omega) \Sigma^r_\beta(\epsilon+\hbar \omega) - G^> (\epsilon) \Sigma^r_\alpha(\epsilon) G^a(\epsilon+\hbar \omega) \Gamma_{\beta}(\epsilon+\hbar \omega) f^e_\beta (\epsilon+\hbar \omega)\right] \\
&-\frac{q^2}{2 \pi \hbar}  \int_{-\infty}^{\infty} d \epsilon \left[ \Gamma_\alpha (\epsilon) G^a(\epsilon) f_\alpha^h (\epsilon) \Gamma_\beta (\epsilon+\hbar \omega) G^a(\epsilon+\hbar \omega) f_\beta^e (\epsilon+\hbar \omega) + G^>(\epsilon-\hbar \omega) \Sigma_\alpha^r (\epsilon-\hbar \omega) G^<(\epsilon) \Sigma_\beta^r (\epsilon)\right].
\end{align}

\end{widetext}

\end{document}